\documentclass[nofootinbib,preprint]{revtex4-1}
\usepackage[T1]{fontenc}
\usepackage{graphicx}
\usepackage{rotating}
\usepackage{bm}        
\usepackage{amssymb}   
\usepackage{bm}
\usepackage{float}
\usepackage{tikz}
\usepackage{diagbox}
\usepackage{braket}
\usepackage{color}
\usepackage{colortbl}
\def\jcp#1#2#3{{\it J.~Chem.~Phys.}~{\bf #1},\ #2\ (#3)}

\def\prl#1#2#3{{\it Phys.~Rev.~Lett.}~{\bf #1},\ #2\ (#3)}

\def\pccp#1#2#3{{\it Phys. Chem. Chem. Phys.}~{\bf #1},\ #2\ (#3)}

\def\jctc#1#2#3{{\it J. Chem. Theor. Comp.}~{\bf #1},\ #2\ (#3)}
\def\ijqc#1#2#3{{\it Int. J. Quant. Chem.}~{\bf #1},\ #2\ (#3)}

\makeatletter

    \def\CT@@do@color{%
      \global\let\CT@do@color\relax
            \@tempdima\wd\z@
            \advance\@tempdima\@tempdimb
            \advance\@tempdima\@tempdimc
    \advance\@tempdimb\tabcolsep
    \advance\@tempdimc\tabcolsep
    \advance\@tempdima2\tabcolsep
            \kern-\@tempdimb
            \leaders\vrule
                    \hskip\@tempdima\@plus  1fill
            \kern-\@tempdimc
            \hskip-\wd\z@ \@plus -1fill }
    \makeatother

\def\k1{k_1}
\def\k2{k_2}
\def\q1{q_1}
\def\q2{q_2}

\def\({\left (}
\def\){\right )}
\def\[{\left [}
\def\]{\right ]}

\newcommand\blfootnote[1]{%
  \begingroup
  \renewcommand\thefootnote{}\footnote{#1}%
  \addtocounter{footnote}{-1}%
  \endgroup
}

\newcommand{\beq}{\begin{equation}}
\newcommand{\eeq}{\end{equation}}

\DeclareMathAlphabet\mathbfcal{OMS}{cmsy}{b}{n}

\begin{document}
\date{\today}
\flushbottom \draft
\title{
Bayesian optimization for inverse problems in time-dependent quantum dynamics
}
\author{Z. Deng$^{a, \ast}$, I. Tutunnikov$^{b, \ast}$, I. Sh. Averbukh$^b$, M. Thachuk$^a$, and R. V. Krems$^{a,c}$}
\affiliation{
$^a$Department of Chemistry, University of British Columbia, Vancouver, B.C. V6T 1Z1, Canada \\
$^b$AMOS and Department of Chemical and Biological Physics, The Weizmann Institute of Science, Rehovot, 7610001, Israel \\
$^c$Quantum Matter Institute, University of British Columbia, Vancouver, BC, Canada V6T 1Z4
}

\blfootnote{$^\ast$ These authors have contributed equally.}

\begin{abstract}
We demonstrate an efficient algorithm for inverse problems in time-dependent quantum dynamics based on feedback loops between Hamiltonian parameters and the solutions of the Schr\"{o}dinger equation.
Our approach formulates the inverse problem as a target vector estimation problem and uses Bayesian surrogate models of the Schr\"{o}dinger equation solutions to direct the optimization of feedback loops. For the surrogate models, we use Gaussian processes with vector outputs and composite kernels built by an iterative algorithm with Bayesian information criterion (BIC) as a kernel selection metric. The outputs of the Gaussian processes are designed to model an observable
simultaneously at different time instances. We show that the use of Gaussian processes with vector outputs and the BIC-directed kernel construction reduce the number of iterations in the feedback loops by, at least,  a factor of 3.
We also demonstrate an application of Bayesian optimization for inverse problems with noisy data. To demonstrate the algorithm, we consider the orientation and alignment of polyatomic molecules SO$_2$ and propylene oxide (PPO) induced by strong laser pulses.  We use simulated time evolutions of the orientation or alignment signals to determine the relevant components of the molecular polarizability tensors to within 1 \% accuracy. We show that, for the five independent components of the polarizability tensor of PPO, this can be achieved with as few as 30 quantum dynamics calculations.

\end{abstract}

\maketitle

\section{Introduction}

The inverse problem in quantum dynamics aims to determine Hamiltonian parameters from experimental observables \cite{ip}.
This goal underlies spectroscopy applications \cite{Levine1975} aiming to determine molecular properties, such as vibrational and rotational constants.
Kraitchman analysis \cite {Kraitchman1953} can be considered an example of an inverse problem that has
been used to determine heavy atom structures of molecules (for a recent example involving
a complex chiral molecule see \cite{Marshall2017}). Electron and x-ray diffraction measurements, which inherently rely on data inversion,
have been applied in molecular physics to probe ultrafast molecular processes
and recover structures of small gas-phase molecules (see, e.g., \cite{Kierspel2020} and
references therein). The Coulomb explosion technique has been improved to allow retrieval of detailed and extremely precise information on molecular
properties from experimental measurements (see e.g. \cite{Pitzer2013, Karamatskos2019} and references therein).
For many quantum systems, it is possible to establish the relationship between the Hamiltonian parameters and observables using
model Hamiltonians and approximations permitting closed-form expressions \cite{le-roy, ip-2} or simplified numerical calculations of observables, which make the inverse problem feasible.
In the present work, we consider an approach to solving inverse problems in quantum dynamics that does not rely on any assumptions about the Hamiltonian and that can be applied to problems, where calculating dynamical observables is extremely time consuming.
This can be used for applications in molecular scattering theory, such as the empirical construction of potential energy surfaces for complex molecular systems, including molecular collision and reaction complexes,
calibration of theoretical approaches, and studies of mechanisms of energy transfer and chemical reactions in molecular collisions.

The present approach is based on a feedback loop, which adjusts Hamiltonian parameters iteratively to bring predictions from the solutions of the Schr\"{o}dinger equation into agreement with experimental observations \cite{BO}. This approach is widely exploited in optimal control experiments aiming to design external field parameters that yield desired quantum dynamics. While optimal control has been successfully applied to many quantum systems, the inverse problem in quantum dynamics remains a significant challenge. Solving the inverse quantum problems by iterative feedback loops is challenging because: (1) each iteration requires the numerical solution of the nuclear Schr\"{o}dinger equation, which is time-consuming; (2) finding optimal Hamiltonian parameters requires global optimization within an unknown range of values; (3) the number of parameters and the dimensionality of the Hamiltonian increase with the complexity of the quantum system, as does the numerical difficulty of solving the  Schr\"{o}dinger equation.

A major thrust of recent work in molecular dynamics research has been to combine traditional simulation methods with machine learning (ML); see, e.g., \cite{bayesian-calibration-1,bayesian-calibration-2, bayesian-calibration-3, bayesian-calibration-4, bayesian-calibration-5, bayesian-calibration-6, bayesian-calibration-7, ML-for-MD-1, ML-for-MD-2, ML-for-MD-3, jie-prl, BML}.
ML offers new tools for solving inverse problems. For example,
it was recently shown that the inverse scattering problem in quantum reaction dynamics can be solved by Bayesian optimization (BO) based on Gaussian processes (GP) as surrogate models of the solutions of the Schr\"{o}dinger equation \cite{BO}.
BO is an efficient, gradient-free optimization approach designed for finding the global extrema of non-convex black-box functions \cite{bo1,bo2}. In this example, the black-box function is the departure of $f(\bm x)$ from a target, with
$f(\bm x)$ representing the dependence of a calculated observable on the Hamiltonian parameters $\bm x$:
\begin{eqnarray}
f(\bm x) \triangleq \left \{ \hat{H}(\bm x) \Psi = E \Psi \rightarrow {\rm Observable}(\bm x) \right \}.
\end{eqnarray}
In \cite{BO}, $\bm x$ represented the potential energy surfaces (PES) determining reactive scattering dynamics.
More generally, the vector $\bm x$ can represent a set of any parameters that determine the Hamiltonian (e.g., PES, non-adiabatic couplings, molecular properties, molecule-field interactions).

 GPs provide an efficient way to explore the dependence of $f$ on $\bm x$ with a small number of quantum calculations.
 This was exploited in \cite{BO} to design the feedback loop

\hspace{4.cm}
\begin{minipage}{0.4\columnwidth}
\centering
 \begin{tikzpicture}[->,scale=.7]
   \node (i) at (90:1cm)  {\small ${\bm x}$};
   \node (j) at (-30:1cm) {\small $f(\bm x)$};
   \node (k) at (210:1cm) {\small $T$};
   \draw (70:1cm)  arc (70:-10:1cm);
   \draw (-50:1cm) arc (-50:-130:1cm);
   \draw (190:1cm) arc (190:110:1cm);

\end{tikzpicture}
\end{minipage} \hspace{4.cm} (2)
\setcounter{equation}{2}

\noindent
where $T$ is a scalar quantity related to an observable. For example, the purpose of feedback loop (2) in \cite{BO} was to obtain the PES $\bm x$ that, when used in a quantum scattering calculation, yielded the energy dependence of the reaction probabilities in agreement with previously published results. To achieve this, $T$ was set to represent the root mean square deviation of $f(\bm x)$ from the reference energy dependence of the reaction probabilities and feedback loop (2) was guided by the minimization of a single parameter $T$. GPs make feedback loop (2) feasible.
This is achieved by building a GP model ${\cal F}(\bm x)$  of $T(\bm x)$ with a small number of quantum calculations. The model ${\cal F}(\bm x)$ is then used by BO to determine how $\bm x$ should be modified in order to minimize $T(\bm x)$ \cite{BO}.

In the present work, we report three significant results.
First, we demonstrate an improved version of BO for inverse problems with time-dependent observables.
We propose an algorithm that represents a time-dependent observable as a vector with $m$ components corresponding to different time instances. This vector is modelled by a GP $\mathbfcal{F}(\bm x)$ with $m$ outputs.
We show that BO based on such GPs converges much faster than algorithm (2) based on GPs ${\cal F}(\bm x)$ with a scalar output.
The vector algorithm can be illustrated by the feedback loop

\hspace{4.cm}
\begin{minipage}{0.4\columnwidth}
\centering
 \begin{tikzpicture}[->,scale=.7]
 \node at (0, 0)   (a) {$\bm x$};
    \node at (3, 0)   (b) {$f(\bm x, t)$};
    \node at (3, -2)  (c)     {$\mathbfcal{F}(\bm x)$};
    \node at (0, -2)  (d)     {$T$};
\draw (a)   -- (b);
\draw (b)   -- (c);
\draw (c)   -- (d);
\draw (d)   -- (a);
\end{tikzpicture}
\end{minipage} \hspace{4.cm} (3)
\setcounter{equation}{3}
where the observable $f$ is calculated at different times $t$ for a given $\bm x$ and the GP model is trained by these calculations to model the observable {\it simultaneously} at different times. The target quantity $T$ is then calculated from $\mathbfcal{F}(\bm x)$ inheriting the properties of the GP with $m$ outputs, which are used to direct BO.
Second, we show that BO can be applied for inverse problems with noisy observables. This is important for applications using raw experimental data for determining the Hamiltonian parameters. Third, we explore the effect of GP model complexity on the efficiency of BO for quantum inverse problems and demonstrate that a model-selection method based on Bayesian information criterion (BIC) can be used to reduce the number of iterations in feedback loop (3).  This model-selection algorithm was previously used to enhance the accuracy of GP models for pattern recognition \cite{extrapolation-1,extrapolation-2}, generalization  \cite{extrapolation-3},  interpolation and extrapolation \cite{extrapolation-3, jun-dai}. Here, we demonstrate that the BIC can also be used to enhance the efficiency of BO. We propose a specific algorithm that identifies the proper GP model complexity before BO is initialized and iterates the feedback loops with the model complexity thus determined.

To demonstrate the algorithms proposed here, we consider the orientation and alignment of polyatomic molecules SO$_2$ and propylene oxide CH$_3$CHCH$_2$O (denoted hereafter as PPO) induced by strong laser pulses.  We use the time dependence of the orientation and/or alignment signal to determine the relevant components of the polarizability tensor of the molecule.
There are several traditional techniques for measuring the average molecular polarizability, including
dielectric constant and refractive index measurements, as well as polarizability anisotropy with the help of
Rayleigh scattering, separately, or in combination with Kerr effect measurements. In some cases, principal
polarizabilities can be recovered by combining the data obtained from two or more independent measurements. For
an in-depth review of these and other modern experimental techniques on electric-dipole polarizabilities, see \cite{Bonin1997}.
The method proposed here may allow measurements of molecular polarizabilities in the gas phase in a single
experimental setup. Moreover, in the case of chiral molecules, our approach allows the determination of the
off-diagonal polarizability tensor components (when expressed in the frame of principal axes of inertia tensor).
The off-diagonal elements determine the relationship between the principal axes of inertia and polarizability,
which is important for various applications (e.g. in molecular dynamics simulations).
We examine the relative importance of the orientation and alignment signals for the inverse problem and show that certain components of the polarizability tensor can be determined from the signal evolution over a short time interval.  Using these examples, we examine the convergence of BO and the accuracy of the Hamiltonian parameters that can be determined with this approach.

\section{Theory}

Feedback loops (2) and (3) require three essential ingredients: numerical integration of the Schr\"{o}dinger equation with a given $\bm x$ to yield $f({\bm x}, t)$, construction of GP models of $f({\bm x},t )$, or some quantity related to $f({\bm x},t )$,  using the results of the quantum calculations as training data, and Bayesian optimization directed by these GP models to find the optimal $\bm x = \bm x^{\rm opt}$.  The quantum problem must be solved at each iteration of the loops. Quantum dynamics calculations are time-consuming and
should be expected to bottleneck the feedback loops. Therefore, it is important to develop an optimization algorithm that converges to $\bm x^{\rm opt}$ with as few iterations as possible.

\subsection{Quantum dynamics calculations}

We consider the asymmetric top molecules treated as rigid tops with three
distinct moments of inertia $(I_{a}<I_{b}<I_{c})$ corresponding to
three principal axes of inertia $a$, $b$ and $c$ \citep{Zare1991,Goldstein2001}.
The interaction with a moderately intense nonresonant short laser
pulse is modeled using the potential (in atomic
units, a.u.) \citep{Lemeshko2013,Krems2018}
\begin{equation}
U=-d_{i}\tilde{E}_{i}-\frac{1}{2}\alpha_{ij}\tilde{E}_{i}\tilde{E}_{j},\label{eq:Int. potential}
\end{equation}
where $d_{i}$ are the components of the molecular dipole moment, $\alpha_{ij}$
are the components of the molecular polarizability and $\tilde{E}_{i}$
are the components of the rapidly oscillating electric field.
The central frequency of an optical laser pulse is of the order
of $\propto10^{15}$ Hz, while the typical rotational frequency of
small molecules is of the order of $h/I\propto10^{12}$
Hz, where $h$ is the Planck's constant and $I$ is the average moment of inertia. Therefore, we average the interaction
energy over the optical cycle, which leads to $U=-\alpha_{ij}E_{i}E_{j}/4$, where $E_i$ are the components of the slowly varying envelope of the laser pulse.
We use the impulsive approximation to treat the interaction of molecules with the laser fields. More details are given in the supplementary information.

We use symmetric top wave functions$\ket{JKM}$, defined in \citep{Zare1991}, to express the Hamiltonian of the asymmetric-top molecule including the interaction term. The Hamiltonian is then transformed to the basis of asymmetric-top eigenfunctions that diagonalizes the kinetic part of the Hamiltonian, allowing for an efficient field-free propagation of wave packets. We aim to explore inverse problems with quantum dynamics calculations for both zero and finite temperatures.
For zero temperature results, we propagate all relevant quantum states in time and compute the resulting expectation values.
At temperatures near 300 K, the number of populated quantum states for molecules such as SO$_2$ or PPO becomes prohibitively large.
Therefore, for all non-zero temperature results, we use a random phase wave functions (RPWF) approach \citep{Kallush2015}. Instead of propagating
each thermally populated state, multiple states  $\ket{\psi_{m}}$ are prepared as mixtures of \emph{all }basis states each
weighted by the Bolzmann factor and assigned a random phase factor, that is
\begin{equation}
\ket{\psi_{m}}=\sum_{n=1}^{n_{\mathrm{max}}}\sqrt{\frac{e^{-\varepsilon_{n}/k_{\mathrm{B}}T}}{Q}}e^{-i\phi_{n}}\ket{n},\label{eq:RPWF-mixture}
\end{equation}
where $\varepsilon_{n}$ is the kinetic energy of the basis state $\ket{n}$,
$k_{\mathrm{B}}$ is the Bolzmann constant, $T$ is the temperature,
$Q$ is the partition function, and $\phi_{n}$ is a random
variable uniformly distributed over
$[0,2\pi]$.  The time evolution of wave packets in Eq. (\ref{eq:RPWF-mixture}) is determined by the time-dependence of the stationary states $|n \rangle$.
A total of
$N$ states (\ref{eq:RPWF-mixture})  is used and the expectation
value of any observable $\hat{O}$ is evaluated as
\begin{equation}
\braket{\hat{O}}\left(t\right)=\frac{1}{N}\sum_{m}\braket{\psi_{m}\left(t\right)|\hat{O}|\psi_{m}\left(t\right)}.\label{eq:RPWF-observable}
\end{equation}
This approach was shown \citep{Kallush2015,Damari2016} to be efficient
for dynamics simulations at elevated temperatures where it converges
with $N \lesssim 100$.

In this study, we consider two molecules (SO$_2$ and PPO) and use three observables: the optical birefringence, the degree of alignment and the orientation factor.
When anisotropic molecules in the gas phase are excited by
a short non-resonant laser pulse, the interaction between the induced
dipole and the electric field tends
to order molecules along the polarization direction of the laser field.
Macroscopically,
molecular alignment leads to measurable optical birefringence of the gas \citep{Faucher2011,Damari2016,Bert2020}. For linear molecules, the
degree of birefringence is determined by the degree of alignment $\braket{\cos^{2}\theta}$, where
$\theta$ is the angle between the molecular axis and the polarization
direction.
For molecules with three distinct principal axes
and three distinct polarizabilities, the optical birefringence is
given by the weighted average of three alignment terms

\begin{eqnarray}
B\left(t\right)\propto\sum_{i=x,y,z}k_{i}\braket{\cos^{2}(\theta_{iZ})},
\label{birefringence}
\end{eqnarray}
where $k_{z}=\alpha^{zx}+\alpha^{zy}$, $k_{x}=-(2\alpha^{zx}+\alpha^{zy})$, $k_{y}=\alpha^{zx}-2\alpha^{zy}$, $\alpha^{zx}=\alpha_{zz}-\alpha_{xx}$, $\alpha^{zy}=\alpha_{zz}-\alpha_{yy}$ and $\theta_{iZ}$ are the angles between the corresponding axes of the rotating molecule fixed frame and the laser pulse polarization
(chosen to be along the laboratory $Z$ axis). The
birefringence depends explicitly on two differences of polarizabilities,
$\alpha^{zx}$ and $\alpha^{zy}$. We use the birefringence (\ref{birefringence}) for the case of alignment of SO$_2$ by a single linearly polarized laser pulse \citep{Stapelfeldt2003,Lemeshko2013,Koch2019}.
In the case of alignment of PPO by a single laser pulse, we use the time dependence of the degree of alignment of the molecular axis $a$ (which is close to the most polarizable axis) with respect to the laboratory $Z$ axis $\braket{\cos^{2}(\theta_{aZ})}$. The supplementary information provides more details.


In the case of orientation of PPO, the observable is the expectation value of the projection of the molecular dipole
vector $\boldsymbol{\mu}$ on the laboratory $Z$ axis, $\mu_{Z}=\boldsymbol{\mu}\cdot\hat{\mathbf{Z}}=\mu\cos(\theta_{\mu Z})$,
where $\mu=|\boldsymbol{\mu}|$ is the magnitude of the dipole moment,
$\hat{\mathbf{Z}}$ is the unit vector along the laboratory $Z$ axis,
and $\theta_{\mu Z}$ is the angle between the molecular dipole moment
and the $Z$ axis. It was recently shown that chiral molecules can be oriented by a pair
of delayed cross-polarized short laser pulses (or more generally by pulses
with twisted polarization). The orientation is in the direction perpendicular to the plane defined by the
two pulses (see \citep{Lin2020} and references therein).
We consider two pulses  in the $XY$ plane, such that the orientation
is along/against the $Z$ axis, and quantify orientation by the dipole orientation factor $\mu \braket{\cos(\theta_{\mu Z})}$ \citep{Babilotte2016,Damari2016}.
The orientation effect stems from  the off-diagonal elements of the polarizability tensor
(in the frame of the principal axes of inertia tensor) $\alpha_{ab},\alpha_{ac}$
and $\alpha_{bc}$, which is a distinct property of chiral molecules.

\subsection{Gaussian process models}

The algorithm described in the next section requires GP models as an intermediate step.
A GP can be considered as a limit of a Bayesian neural network with an infinite number of hidden nodes \cite{BML, neal}. The inputs to the GP are $N$ independent variables, collectively denoted by the vector $\bm x = \left [ x_1, ..., x_N \right ]^\top$.
The output  of a GP is a scalar function $y(\bm x)$.
The purpose is to model an ensemble of $n$ data points $\bm y = \left [Y_1, ..., Y_n \right ]^\top$ located at $\left [ \bm x_1, ..., \bm x_n \right ]^\top$ of the $N$-dimensional variable space. It is assumed that these data points can be described as  $f(\bm x) + \varepsilon$ , where $f(\bm x)$ is some function and $\varepsilon$ is Gaussian-distributed noise with variance $\sigma^2$.

 At any $\bm x$, there is a normal distribution $P(y)$ of values $y$.
When a GP is trained,  $P(y)$ is conditioned by the $n$ data points $\bm y$ at $\left [ \bm x_1, ..., \bm x_n \right ]^\top$. The mean and variance of this conditional distribution at an arbitrary point $\bm x_\ast$ are given by \cite{BML,gp-book}
\begin{eqnarray}
\mu_\ast = \bm k_\ast^\top (\bm K + \sigma^2 \bm I)^{-1} \bm y,
\label{GP-mean}
\end{eqnarray}
\begin{eqnarray}
\sigma_\ast = k(\bm x_\ast, \bm x_\ast)  - \bm k_\ast^\top  (\bm K + \sigma^2 \bm I)^{-1} \bm k_\ast,
\label{GP-variance}
\end{eqnarray}
where $\bm k_\ast$ is a vector with $n$ entries $k(\bm x_\ast, \bm x_i)$ and $\bm K$ is an $n \times n$ matrix with elements $k(\bm x_i, \bm x_j)$.
Eq. (\ref{GP-mean}) is used to predict the value of $f(\bm x)$ at $\bm x_\ast$.
The quantities $k(\bm x_i, \bm x_j)$ are the kernels, which, with a particular choice of the GP prior \cite{BML,gp-book}, represent the covariance of the normal distributions of $y$ at $\bm x_i$ and at $\bm x_j$.

A particular mathematical form of the kernel function $k({\bm x}, {{\bm x}'})$ defines a GP model. The choice of this function is not unique.
Unless specified otherwise (c.f., Section III.B), we use  the functional form  \cite{mitchell1990existence, cressie1993statistics, stein1999interpolation}
\begin{eqnarray}
 k({\bm x}, {{\bm x}'})  = \frac{2^{1-v}}{\Gamma(v)}\left( \sqrt{2v}r({\bm x}, {{\bm x}'}) \right )^v \mathcal{K}_v \left ( \sqrt{2v}r({\bm x}, {{\bm x}'})\right )~~~~
\label{eqn:k_MAT}
\end{eqnarray}
where $r^2({\bm x}, {{\bm x}'}) = ({\bm x}- {{\bm x}'})^\top \times {\bm M} \times ({\bm x}-{{\bm x}'})$ and ${\bm M}$ is a diagonal matrix with $N$ parameters, one parameter for each dimension of ${\bm x}$,
 $\mathcal{K}_v$ is the modified Bessel function, $\Gamma$ is the
Gamma function, and $v=3/2$.
This function is often referred to as the Mat\'{e}rn function.
The parameters of the kernel function are found by maximizing the logarithm of the marginal likelihood \cite{BML,gp-book}
\begin{eqnarray}
\log {\cal L} = -\frac{1}{2}{\bm y}^\top  {\bm K} ^{-1}{\bm y} - \frac{1}{2}\log |\bm K | - \frac{n}{2} \log 2\pi.
\label{log-likelihood-explicit}
\end{eqnarray}

\subsection{Bayesian optimization with scalar and multiple-output GPs}

There are six independent matrix elements that determine the polarizability tensor of a general polyatomic molecule in the frame of the principal axes of inertia tensor $\alpha_{aa}, \alpha_{bb}, \alpha_{cc}, \alpha_{ab}, \alpha_{ac}, \alpha_{bc}$.
Depending on the symmetry of the molecule, some of the polarizability tensor matrix elements vanish. In addition, different observables are determined by different parts of the polarizability tensor.
We define the vector $\bm x$ as comprising the minimum number of independent parameters of the polarizability tensor determining an observable.
This is the maximum amount of information that can be inferred by solving the inverse problem.
For example, the laser-field alignment of the planar molecule SO$_2$ is determined by two parameters: $\alpha^{ab} =  \alpha_{aa} - \alpha_{bb} $ and $\alpha^{ac} = \alpha_{aa} - \alpha_{cc} $.
For this case, we define $\bm x$ as $\bm x = \left [ \alpha^{ab}, \alpha^{ac} \right ]^\top$.
The orientation of PPO by laser pulses with twisted polarization is determined by five parameters: $\alpha^{ab}$, $\alpha^{ac}$, $\alpha_{ab}$, $\alpha_{ac}$ and $\alpha_{bc}$.
For this case, we define $\bm x$ as the five-dimensional vector $\bm x = \[ \alpha^{ab}, \alpha^{ac}, \alpha_{ab}, \alpha_{ac}, \alpha_{bc} \right ]^\top$.
Given $\bm x$, quantum dynamics calculations described in Section I.A produce the observable (the birefringence signal probing alignment or the orientation factor characterizing orientation) as a function of time $t$.
This observable is hereafter denoted $f(\bm x, t)$.

The purpose of BO is to find $\bm x = \bm x^{\rm opt}$ that leads to $f(\bm x^{\rm opt}, t)$ in agreement with some reference function $f_{\rm ref}(t)$ giving the time dependence of the observable. The departure of $f(\bm x, t)$ from $f_{\rm ref}(t)$ can be quantified by the root mean square error (RMSE)
\begin{eqnarray}
{\cal E}(\bm x) = \frac{1}{m} \left \{ \sum_{i}^m \left [  f(\bm x, t_i) - f_{\rm ref}(t_i)   \right ]^2 \right \}^{1/2}
\label{target-rmse}
\end{eqnarray}
with $t_i$ representing discretized points of time along a chosen time interval. The vector $\bm x^{\rm opt}$ corresponds to the minimum of ${\cal E}(\bm x)$.
We treat the quantities $f(\bm x, t)$ and $f_{\rm ref}(t)$ as vectors $\bm f$ and $\bm f_{\rm ref}$ with $m$ entries corresponding to discretized time values $t_i$.
Feedback loops (2) and (3) provide two different algorithms to find the minimum of ${\cal E}(\bm x)$.

For loop (2), we define $T(\bm x)$ as
\begin{eqnarray}
T(\bm x) = - \arccos \left (  \frac{{\bm f} \cdot {\bm f}_{\rm ref}}{|{\bm f}| |{\bm f}_{\rm ref}| }   \right ),
\label{cos-sim}
\end{eqnarray}
which quantifies the similarity between the vectors.
Loop (2) uses a scalar, single-output GP model of $T(\bm x)$ thus defined.
The algorithm starts by evaluating $T(\bm x)$ at a small number of randomly chosen values $\left [ \bm x_1, ..., \bm x_n \right ]^\top$. The results of quantum dynamic calculations for these $\bm x$ are used to build a GP model of
$T(\bm x)$. The subsequent calculation of $T(\bm x)$ is performed at the value of $\bm x = \tilde {\bm x}$ that corresponds to the maximum of the acquisition function defined as
\begin{eqnarray}
\alpha(\bm x) = \mu(\bm x) + \kappa \sigma (\bm x),
\label{aq-simple}
\end{eqnarray}
where $\mu$ and $\sigma$ are given by Eqs. (\ref{GP-mean}) and (\ref{GP-variance}), respectively, and $\kappa$ is a fixed parameter.
The value of $T(\tilde {\bm x})$ is added to the training points for the GP to generate a new GP with improved values of $\mu$ and $\sigma$ and the process is iterated.
The acquisition function defined by Eq. (\ref{aq-simple}) provides a balance between exploration of the entire variable parameter space, driven by the second term, and exploitation of a promising part of the parameter space, driven by the maximization of the first term.
As the number of points in the $\bm x$ space increases, $\sigma(\bm x)$ decreases and the maximum of the acquisition function tends to the maximum of  $T(\bm x)$ in Eq. (\ref{cos-sim}).
Note that $T(\bm x)$ in loop (2) can be chosen as ${\cal E}(\bm x)$, as was done in the previous work \cite{BO}. In the present work, we found that the choice of the scalar function (\ref{cos-sim}) instead of ${\cal E}(\bm x)$ reduces the number of BO iterations.


Loop (3) minimizes the quadratic Euclidean distance, or the square of the 2-norm, between $\bm f$ and $\bm f_{\rm ref}$, that is $T(\bm x) = | {\bm f}(\bm x) -  {\bm f}_{\rm ref} |^2$. Instead of using one GP model to represent $T(\bm x)$, at each iteration, algorithm (3)
uses $m$ GP models  trained to represent the $m$ components of the vector $\bm f$, as described in \cite{norm-minimization}. The $m$ GP models are trained simultaneously, assuming the same ($\bm x, \bm x'$) covariance for different $t_i$. This leads to a more complex form of the acquisition function. Whereas in algorithm (2), $T(\bm x)$ is directly represented by a GP, the square of the 2-norm in algorithm (3) follows a non-central Chi-squared distribution with $m$ degrees of freedom. As was shown in \cite{norm-minimization}, this approach
is more efficient for minimization of distances than the method using Eq. (\ref{aq-simple}), because it uses the individual vector components for training GPs and leads to an acquisition function based on a non-symmetric distribution better suited for non-negative functions. To obtain the acquisition function,
it is necessary to transform the means and uncertainties of the $m$ GPs into quantities characterizing the Chi-squared distribution and then map these quantities onto the parameters of the acquisition function. This was done in \cite{norm-minimization}, where the authors used the results of \cite{chi2-to-Gaussian} to represent the Chi-squared distribution by an approximate Gaussian distribution. This yields the following result for the  acquisition function suitable for the minimization of the 2-norm \cite{norm-minimization}:
\begin{eqnarray}
\alpha (\bm x)=-\sqrt[l]{\delta-\kappa\rho} (m+\lambda)\gamma^2,
\label{aq-vector}
\end{eqnarray}
where
\begin{eqnarray}
\lambda=\gamma^{-2}\sum_{i=1}^m \left [ \mu_i(\bm x)-{\bm f}_{{\rm ref}, i} \right ]^2,
\end{eqnarray}
\begin{eqnarray}
\gamma=\sqrt{\frac{1}{m}\sum_{i=1}^m\sigma_i^2(\bm x)}
\end{eqnarray}
$\mu_i$ and $\sigma_i$ are the mean and the variance of the $i$th output of the GP with $m$ outputs,
\begin{eqnarray}
l=1-\frac{r_1r_3}{3r_2^2},
\end{eqnarray}
with
\begin{eqnarray}
r_i=2^{i-1}(i-1)!(m+i\lambda),
\end{eqnarray}
\begin{eqnarray}
\rho = \frac{l r_2^2}{r_1}\left ( 1 - \frac{(1-l)(1-3l)}{4 r_1^2}r_2       \right ),
\end{eqnarray}
and
\begin{eqnarray}
\delta=1+l(l+1)\left (\frac{r_2}{2r_1^2}-(2-l)(1-3l)\frac{r_2^2}{8r_1^4} \right ).
\end{eqnarray}
The parameter $\kappa$ is used to determine the balance between exploration and exploitation in the optimization algorithm. In the present work, $\kappa$ was set to $0.8$ in both Eq. (\ref{aq-simple}) and (\ref{aq-vector}).

\begin{figure}
\includegraphics[scale=0.35]{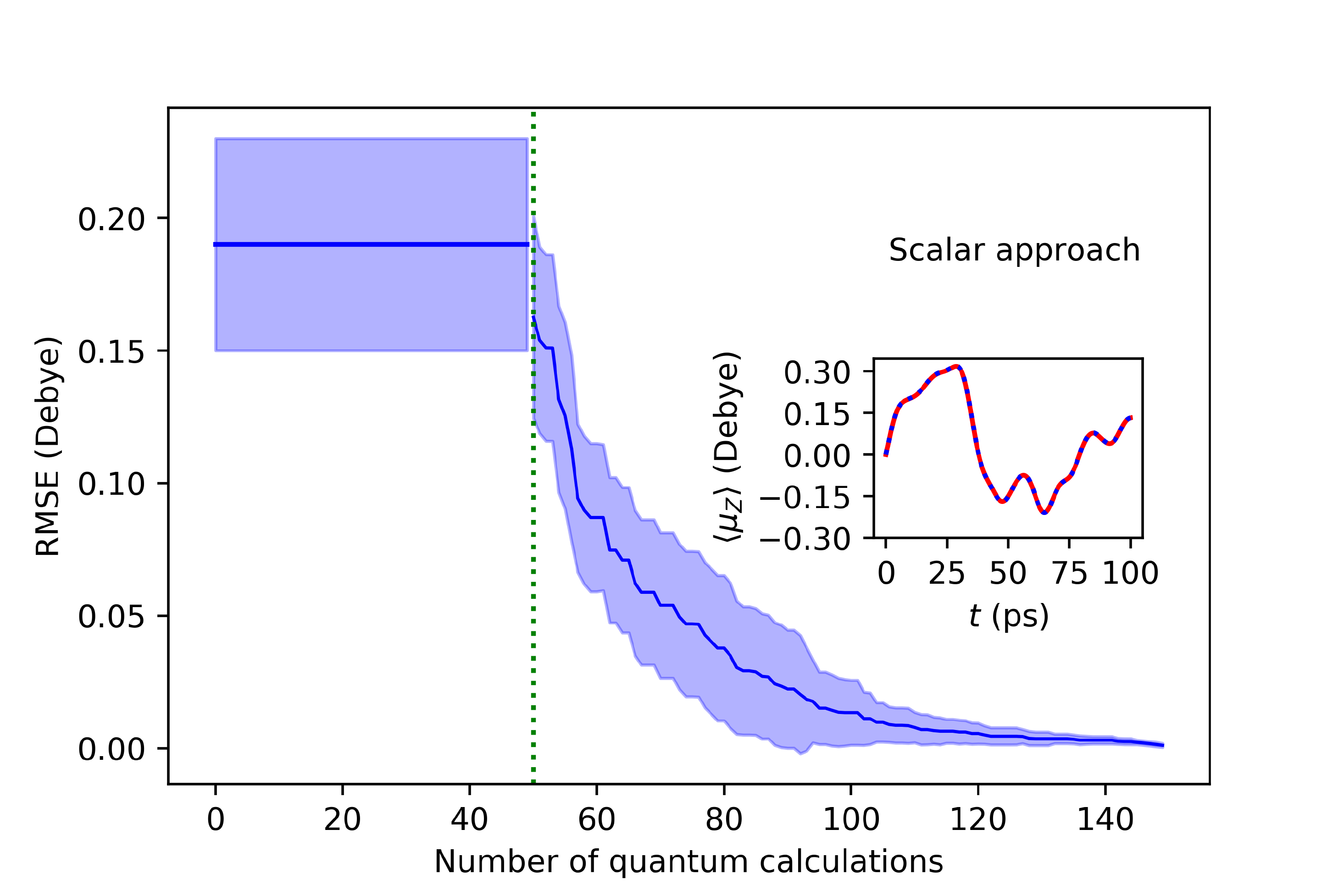}
\includegraphics[scale=0.35]{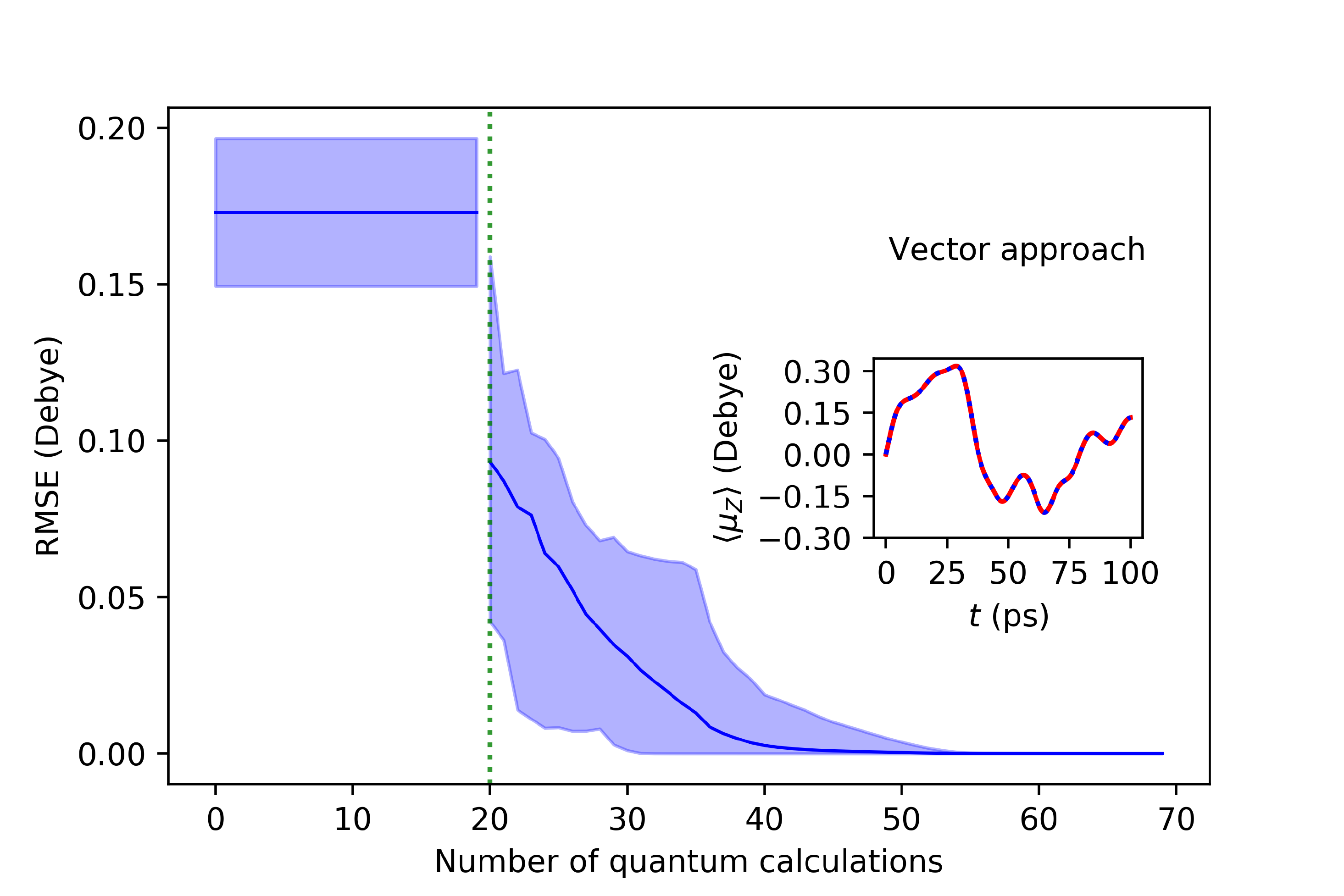}
\caption{The results of BO of the parameters $\alpha^{ab}$, $\alpha^{ac}$, $\alpha_{ab}$, $\alpha_{ac}$ and $\alpha_{bc}$ of PPO using the orientation factor as the reference signal. The vertical dotted line marks the beginning of BO.
For illustration purposes, the optimization was repeated 20 times using different initial conditions. The shaded area to the left of the dotted line shows the standard deviation of RMSE (\ref{target-rmse}) resulting from the distribution of these 20 calculations.
The shaded area to the right of the vertical dotted line spans the range of RMSE from the minimum to maximum values. The solid curve is the average of 20 calculations.
Upper panel: the scalar GP algorithm (2); Lower panel: algorithm (3) based on the multiple-output GP with a complex kernel represented by a linear combination of a Mat\'{e}rn and RQ functions (see Section III.B for details).
Insets: the solid curve shows the reference signal $f_{\rm ref}(t)$ used for BO and the broken curves -- the final result computed with the optimized parameters of the polarizability tensor after 90 (upper panel) and 20 (lower panel) iterations. }
\label{figure-1}
\end{figure}

\section{Results}

The present work considers two molecules: SO$_2$ and propylene oxide (PPO).
The laser-field alignment of the planar molecule SO$_2$ is determined by two parameters: $\alpha^{ab} =  \alpha_{aa} - \alpha_{bb} $ and $\alpha^{ac} = \alpha_{aa} - \alpha_{cc} $. The alignment and orientation of PPO is determined by five parameters \cite{ilia-paper}: $\alpha^{ab}$, $\alpha^{ac}$, $\alpha_{ab}$, $\alpha_{ac}$ and $\alpha_{bc}$. Our goal is to determine the values of these parameters based on a reference signal $f_{\rm ref}(t)$.

For the reference signal $f_{\rm ref}(t)$, we use the birefringence (for SO$_2$) and the alignment and orientation factors (for PPO) computed with the most accurately known theoretical values of the polarizability tensors (see the supplementary information for more details).
Unless otherwise specified, the reference curves are the theoretical calculations at zero temperature. For the calculations at zero temperature, the molecules are assumed to be excited by a pulse with the duration 20 fs and the field intensity $5 \times 10^{13}$ W/cm$^2$.
We also explore the effect of averaging over quantum states required to simulate the experiments at finite temperature.
For the calculations at finite temperature, the pulse duration was chosen to be 100 fs and the field intensity $5 \times 10^{13}$ W/cm$^2$.
The initial optimization was performed blindly with one co-author producing the reference results and another performing optimization without information on the reference polarizability tensor parameters.

In the present section, we consider noiseless functions $f_{\rm ref}(t)$. In the absence of noise, $\sigma$ in Eqs. (\ref{GP-mean}) and (\ref{GP-variance}) must be set to zero.
The effect of noise is considered in Section II.A. We refer to algorithm (2) as the `scalar' algorithm and (3) as the `vector' algorithm.
For all calculations presented here, we discretize the time variable to represent $f(\bm x, t)$ and $f_{\rm ref}(t)$ by 400 equally spaced points in the indicated time interval beginning after the laser pulse (alignment) or the second laser pulse (orientation).

Figure \ref{figure-1} illustrates the main features and convergence of BO. The algorithm illustrated in Figure 1 begins with a fixed number (50 in the upper panel and 20 in the lower panel) of quantum dynamics calculations at different values of $\bm x = \[\alpha^{ab}, \alpha^{ac}, \alpha_{ab}, \alpha_{ac},\alpha_{bc} \right]^\top$ for PPO. The values of these parameters are chosen using the Latin hypercube spacing algorithm to avoid  clustering in this five-dimensional space \cite{LHS,jie-pes}.
The reference function $f_{\rm ref}(t)$ is the expectation value of the $Z$-component of the molecular dipole moment shown in the insets of Figure \ref{figure-1}. The deviation of the calculated signal from the reference curve is quantified by the value of RMSE (\ref{target-rmse}). The final time-dependence of $\braket{\mu_Z}$ calculated with the optimized parameters of the polarizability signal is shown by the broken curves in the insets of Figure \ref{figure-1}.

BO begins at the vertical dotted lines of Figure \ref{figure-1}. To illustrate the efficacy and convergence of BO, we repeat BO with 20 different sets of the initial quantum calculations. The shaded area to the left of the dotted line shows the standard deviation of RMSE (\ref{target-rmse}) resulting from the distribution of these 20 sets. Each of the initial conditions leads to a different BO trajectory, bringing RMSE to zero. The shaded area to the right of the vertical dotted line spans the trajectories from the minimum to maximum values of RMSE. Figure 1 shows that all BO trajectories, regardless of the initial conditions, converge to zero RMSE, and that the convergence is monotonous. The results in the upper panel of Figure \ref{figure-1} were obtained using the scalar approach (2) and the kernel (\ref{eqn:k_MAT}). The subsequent discussion explores how BO can be accelerated by employing feedback loop (3) and modifying the kernel complexity, leading to the results in the lower panel of Figure  \ref{figure-1}.

The numerical results of BO are summarized in Tables I - III. The variable space was set to span $\alpha^{ab}  \in \left [ -30, +30 \right ]$ and $\alpha^{ac}  \in \left [ -30, +30 \right ]$ for SO$_2$; and
$\alpha^{ab}  \in \left [ -40, 40 \right ]$, $\alpha^{ac}  \in \left [ -40, 40 \right ]$, $\alpha_{ab} \in \left [0, 5 \right ]$, $\alpha_{ac} \in \left [0, 5 \right ]$ and $\alpha_{bc} \in \left [0, 5 \right ]$ for PPO, all in atomic units. The observation of these results leads to the following conclusions:

\begin{itemize}

\item The vector approach (3) converges much faster than the scalar approach (2). Note that the kernels of the GP models in these calculations are represented by the same function (\ref{eqn:k_MAT}) so the acceleration of the BO convergence is due to the difference in the algorithms.

\item Both approaches can identify the five parameters of the polarizability  tensor for PPO with accuracy better than 1 \%.

\item The convergence of the calculations based on the alignment and orientation signals is different.

\end{itemize}

\begin{figure}
\includegraphics[scale=0.4]{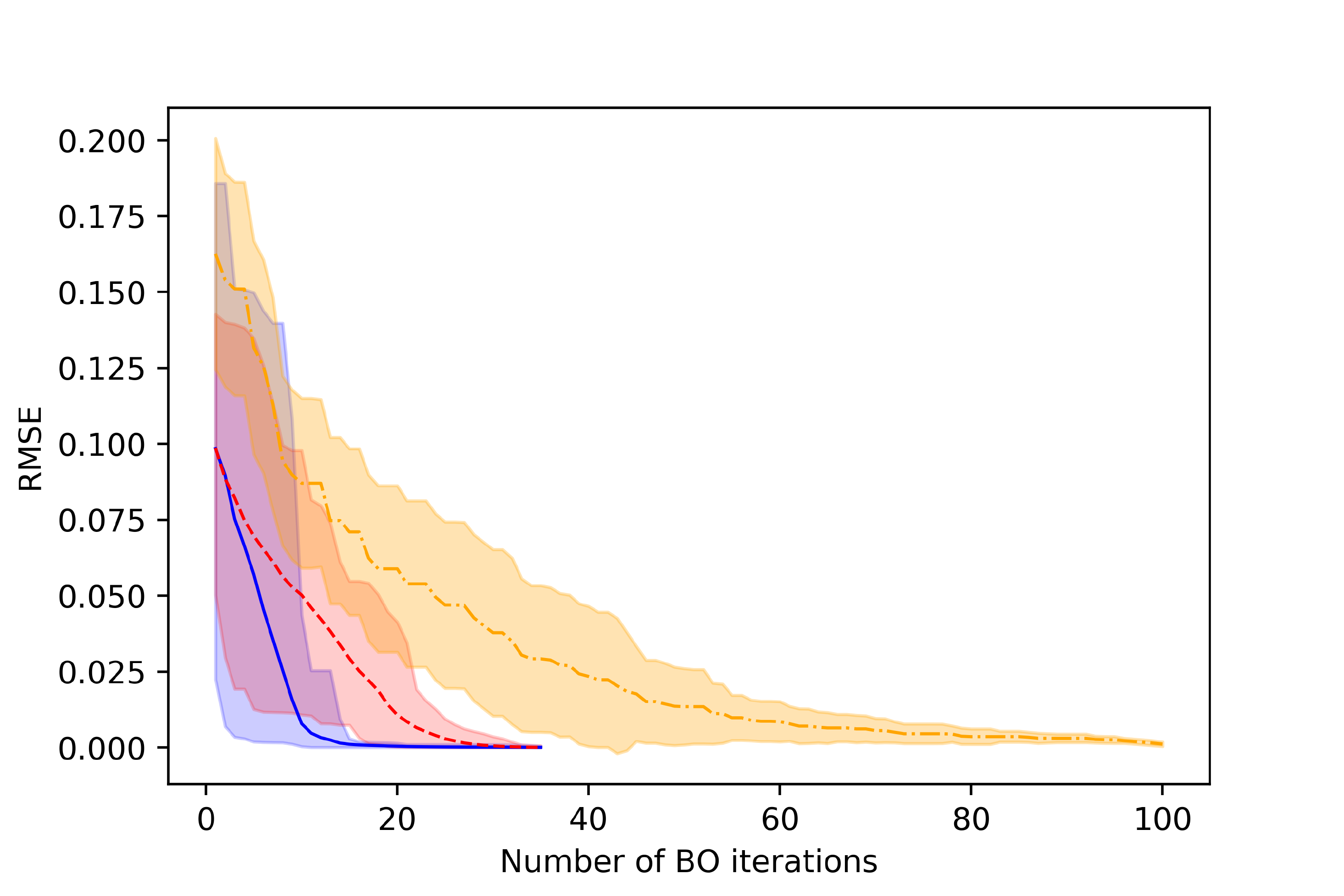}
\caption{Convergence of BO iterations for the 5-dimensional case of PPO: dot-dashed curve (orange) -- scalar BO based on the orientation signal; dashed curve (red) -- vector BO based on the orientation signal; solid curve (blue) -- vector BO based on the alignment signal.
The shaded areas span the range from the minimum to maximum values of RMSE in the set of 20 calculations with different initial conditions. BO is initialized by 50 quantum calculations as in Figure \ref{figure-1}. The RMSE is dimensionless for the alignment and in units of Debye for the orientation.
}
\label{figure-2}
\end{figure}

  Figure \ref{figure-2} elucidates the last observation and illustrates the relative convergence of the scalar and vector algorithms.
  It is clear from the figure that the vector approach reduces the number of BO iterations required at least by a factor of 2.
  The alignment reference signal appears to constrain the polarizability tensor parameters more effectively, resulting in a further reduction of BO iterations required for full convergence. It is important to note that, in practice,
  it is not necessary to repeat BO with different initial conditions. However, it may be advisable to perform calculations with, at least two, drastically different initial conditions to quantify convergence. These calculations are completely independent and can be performed in parallel.
  If quantum dynamics calculations are extremely time consuming, the best approach may be to begin multiple BO calculations with drastically different initial conditions and harvest the results from the BO trajectory approaching the target most steeply.

\begin{table}[h!]
\centering
\begin{tabular}{ c|c|c }
 \hline \hline
 SO$_2$& $\alpha^{ab}$&$\alpha^{ac}$ \\
 \hline
\rowcolor{pink} Reference & 10.46 & 12.62\\
 \hline
 10 ps (scalar) & 10.47 & 12.47 \\
 \hline
 100 ps (scalar) & 10.52 & 12.50 \\
 \hline
 10 ps (vector) & 10.45 & 12.60 \\
 \hline \hline
\end{tabular}
\caption{The polarizability parameters (in a.u.) of SO$_2$ determined by BO with 50 iterations (scalar) and 30 iterations (vector) using the time dependence of birefringence over the indicated time interval. The results shown are averaged over 20 instances of initial conditions for BO. Each initial condition is based on 50 quantum calculations with  $\alpha^{ab}$ and $\alpha^{ac}$ chosen randomly using Latin hypercube sampling.
}
\end{table}

\begin{table}[h]
\centering
\begin{tabular}{ c|c|c|c|c|c }
 \hline \hline
PPO & $\alpha^{ab}$&$\alpha^{ac}$&$\alpha_{ab}$&$\alpha_{ac}$&$\alpha_{bc}$ \\
 \hline
 \rowcolor{pink} Reference & 7.67 & 7.76 & 2.56 & 0.85 & 0.65\\
 \hline
  \multicolumn{6}{c}{Alignment}\\
 \hline
 100 ps (scalar) & 7.461 & 7.937 & 2.581 & 0.796 & 0.560 \\
 \hline
 1000 ps (scalar) & 7.816 & 7.609 & 2.558 & 0.880 & 0.620 \\
 \hline
 100 ps (vector) & 7.669 & 7.761 & 2.560 & 0.850 & 0.649 \\
 \hline
 \multicolumn{6}{ c}{Orientation}\\
 \hline
 100 ps (scalar) & 7.723 &7.753 &2.571 & 0.854& 0.655\\
 \hline
 1000 ps (scalar) & 7.677 & 7.775 & 2.568 & 0.852 & 0.653 \\
 \hline
 100 ps (vector) & 7.666 & 7.760 & 2.560 & 0.850 & 0.649 \\
 \hline \hline
\end{tabular}
\caption{
The polarizability parameters (in a.u.) of PPO determined by BO with 100 iterations (scalar) and 50 iterations (vector) using the alignment (for alignment) and orientation (for orientation) factors over the indicated time interval. The results shown are averaged over 20 instances of initial conditions for BO. Each initial condition is based on 50 quantum calculations with the polarizability parameters chosen randomly using Latin hypercube sampling.
}
\end{table}

\begin{table}[h]
\centering
\begin{tabular}{ c | c|c|c|c|c|c }
 \hline \hline
$N_i$ & PPO & $\alpha^{ab}$&$\alpha^{ac}$&$\alpha_{ab}$&$\alpha_{ac}$&$\alpha_{bc}$ \\
 \hline
 \rowcolor{pink} & Reference & 7.67 & 7.76 & 2.56 & 0.85 & 0.65\\
 \hline
   \multicolumn{7}{c}{Alignment}\\
 \hline

 50 & Average  & 7.669 & 7.761 & 2.560 & 0.850 & 0.649\\
 \hline
50 &  SD & 0.008 & 0.007 & 0.0002 & 0.0006 & 0.002\\
 \hline
30 &  Average & 7.671 & 7.762 & 2.560 & 0.851 & 0.614 \\
 \hline
30 & SD  & 0.145 & 0.119 & 0.002 & 0.006 & 0.150 \\
 \hline
10  & Average  & 17.13 & 16.05 & 3.65& 2.39 & 1.43 \\
 \hline
 10 & SD  & 11.81 & 10.10 & 1.20 & 1.93 & 1.28 \\
 \hline

 \multicolumn{7}{c}{Orientation}\\
 \hline
 50 & Average  & 7.666& 7.760& 2.560& 0.850& 0.649\\
 \hline
 50 & SD & 0.003 & 0.004 & 0.002 & 0.0005 & 0.0009\\
 \hline
 30 & Average  & 7.670 & 7.760 & 2.560 & 0.851 & 0.614 \\
 \hline
 30 & SD  & 0.004 & 0.003 & 0.002 & 0.006 & 0.005\\
 \hline
 10 & Average  & 7.938 & 4.125 & 2.389 & 1.574 & 1.402 \\
 \hline
  10 & SD  & 17.62 & 10.09 & 1.25 & 1.45 & 1.59 \\
 \hline \hline
\end{tabular}
\caption{
The polarizability parameters (in a.u.) of PPO determined by BO after 35 iterations using the same reference signal as in Table II.
$N_i$ represents the number of quantum calculations before BO is initiated (vertical dotted line in Figure 1).
The averages and standard deviations (SD) are computed using 20 initial conditions.
}
\end{table}

\clearpage
\newpage

\begin{figure}
\centering
\includegraphics[scale=0.5]{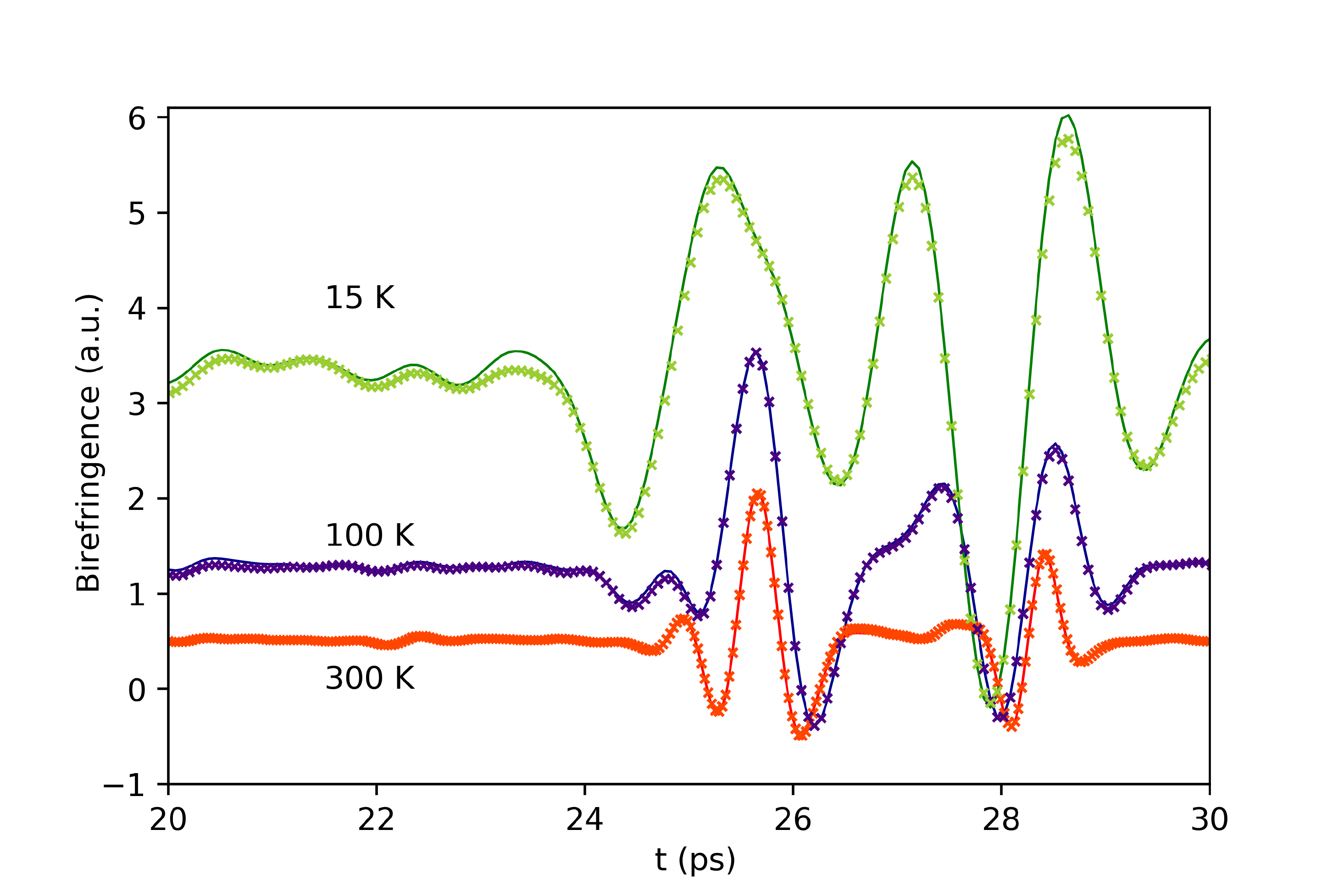}
\caption{
Time dependence of the birefringence of a gas of SO$_2$ at three different temperatures ($T=15, 100$ and $300$ K): solid curves --  calculation with the reference values of the polarizability tensor parameters; symbols -- calculations with the polarizability tensor parameters determined by BO (c.f., Table IV).
}
\end{figure}

\begin{table}[h]
\centering
\begin{tabular}{ c | c|c|c }
 \hline \hline
Temperature &  & $\alpha^{ba}$&$\alpha^{bc}$ \\
 \hline
 \rowcolor{pink} Reference &  & -10.46 & 2.16 \\
 \hline
 $15$ K & Average & -10.46 & 2.16 \\
& SD  & 0.092 & 0.085 \\
 \hline
 $ 100$ K & Average & -10.47 & 2.15  \\
 & SD & 0.038 & 0.087 \\
 \hline
  $ 300$ K & Average & -10.47 & 2.15 \\
 & SD& 0.041 & 0.073 \\
 \hline
 \hline \hline
\end{tabular}
\caption{
The polarizability parameters (in a.u.) of SO$_2$ determined by BO (vector approach) with 20 iterations using the birefringence signal over 100 ps at three different temperatures. The results are averaged over 20 instances of initial conditions for BO, leading to the distributions of values with the standard deviations (SD) listed.  Each initial condition is based on 20 quantum calculations with the polarizability parameters chosen randomly using Latin hypercube sampling.}
\end{table}

Figure 3 presents the results of BO using the reference curves computed at finite temperatures. The quantum dynamics calculations are performed using the RPWF approach with $N = 40$, as described in Section II.A.
The polarizability tensor parameters determined from the reference signals shown in Figure 3 are given in Table IV. To illustrate the stability of the predictions, we repeated the calculations with 20 initial conditions, leading to the spreads of the polarizability tensor parameters
characterized by the standard deviations listed in Table IV. The results illustrate that the averaging over quantum states required for the finite temperature simulations does not spoil the accuracy of BO and does not affect the convergence of BO.

We would like to briefly discuss several complications that may arise in recovering the
polarizabilities from experimental signals. Most significantly, in most experiments, the
pulse parameters are not known with sufficient precision. In such cases, the
polarizabilities cannot be directly recovered, because the Hamiltonian involves products of
the pulse intensity and polarizability (see Eq. 4). Moreover, experimental data (e.g.
birefringence signal) are often systematically scaled and shifted. We suggest that an auxiliary experiment
may, in principle, allow one to resolve these issues. In this experiment, the measurement
is done on simple molecules with precisely known polarizability. Using the methods
described here, the pulse parameters, signal scaling, and signal shift can
be optimized by including them in the ${\bm x}$ vector, while keeping the polarizability tensor elements fixed to the accurate values.
Next, the main experiment with the complex molecule of interest is carried out under the
same conditions. At this point, the procedure continues as described here, i.e. BO is
performed with the ${\bm x}$ vector including the polarizabilities of the complex molecule, while
the pulse parameters, and signal scaling/shift are kept fixed at the values found in the
auxiliary experiment.

Depending on the experimental conditions, physical models can be extended to better
describe both the auxiliary and the main experiments. For example, the models can include
averaging over the spatial intensity profile of the laser beam, molecular centrifugal
distortion, hyperfine coupling effects (see, e.g., \cite{hf}), etc.
As in the case of polarizabilities, the additional parameters can be added to the $\bm x$ vector.

\subsection{Effect of noise in reference signal}

Experimental data generally come with noise. GPs are particularly well suited to describe noisy data, as Gaussian noise is naturally built into the construction of the likelihood function \cite{gp-book, BML}. The variance of the noise $\sigma$ can be treated as a variable parameter when training GPs.
BO based on GPs can thus readily account for the experimental noise in the reference signal. To demonstrate this, we modulate the reference birefringence curve to include Gaussian noise in the time domain.
Specifically, we generate a random distribution $f_{\rm ref + noise}(t) = f_{\rm ref}(t) + {\cal N}({\cal M}, \gamma \times \Sigma )$, where $\cal M$ and $\Sigma$ are the mean and the variance of the 400 points in $f_{\rm ref}(t)$, $\cal N$ denotes a normal distribution and $\gamma$ is
a variable percentage factor. For  the signal with 10 \% noise, $\gamma = 0.1$.
The values $f_{\rm ref + noise}(t)$ thus generated are then used as the reference signal for BO.

Table V summarizes the results for PPO based on the reference signal (the orientation factor) with different amounts of noise. As follows from Table V,  5 \% noise allows the determination of the diagonal elements of the polarizability tensor to within 3.4 \% and the small off-diagonal
elements to within 7.5 \%.


Figure 4 shows the results for PPO using the degree of alignment curve with 10 \% noise as the reference signal.
The figure illustrates that the signal calculated with the optimized polarizability parameters is very close to the mean of the reference signal. The diagonal elements of the polarizability tensor as well as $\alpha_{ac}$ are constrained to better than 7 \%,
while the off-diagonal element $\alpha_{bc}$ deviates from the reference by a factor of 2. This suggests that the degree of alignment is much less sensitive to $\alpha_{bc}$ than the other polarizability tensor parameters.
We note that typical experimental data have much lower noise levels.

\begin{table}[h]
\centering
\begin{tabular}{ c|c|c|c|c|c }
 \hline \hline
PPO & $\alpha^{ab}$&$\alpha^{ac}$&$\alpha_{ab}$&$\alpha_{ac}$&$\alpha_{bc}$ \\
 \hline
 \rowcolor{pink} Reference & 7.67 & 7.76 & 2.56 & 0.85 & 0.65\\
 \hline
 \multicolumn{6}{ c}{Orientation}\\
 \hline
 1\% noise & 7.637 &7.799 &2.577 & 0.862& 0.650\\
 \hline
 SD/\% & 2.796 &1.576 &3.103 &4.001 & 4.221\\
 \hline
 3\% noise & 7.656 & 7.750 & 2.580 & 0.844 & 0.662 \\
 \hline
 SD/\% &2.267 &1.765 &4.212 &4.903 &6.475 \\
 \hline
 5\% noise & 7.678 & 7.790 & 2.592 & 0.857 & 0.656 \\
 \hline
 SD/\% & 3.380&3.244 & 5.112&4.426 & 7.492\\
 \hline
 \hline \hline
\end{tabular}
\caption{
The polarizability parameters (in a.u.) of PPO determined by BO  (vector approach) with 50 iterations using the orientation factor over 100 ps of the reference signal including Gaussian noise with magnitude 1\%, 3\%, and 5\%.
The results shown are averaged over 20 instances of initial conditions for BO. The standard deviation (SD) is in percentage with respect to the average values.
Each initial condition is based on 50 quantum calculations with the polarizability parameters chosen randomly using Latin hypercube sampling.
}
\end{table}

\begin{figure}
\centering
\includegraphics[scale=0.5]{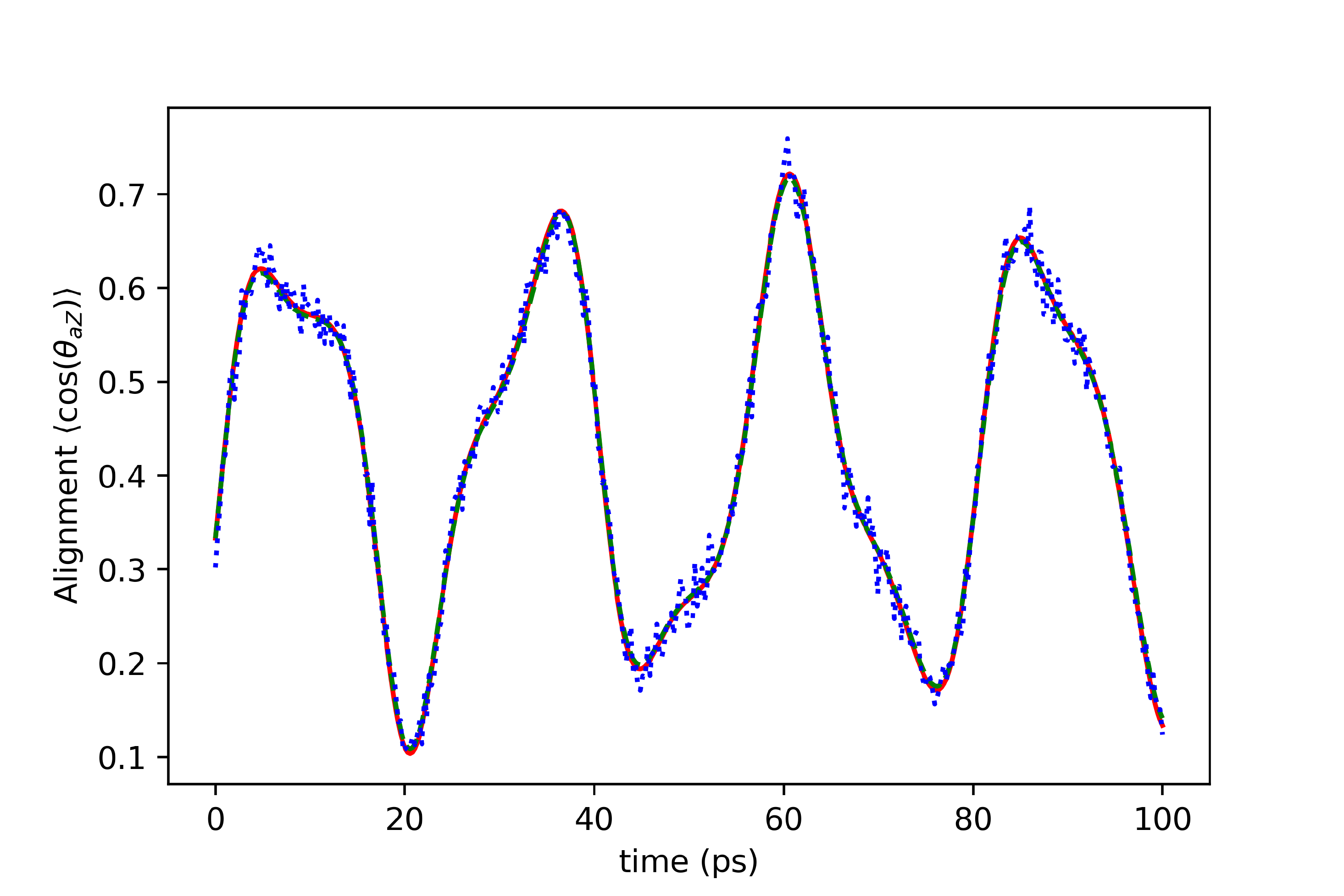}
\caption{Comparison of the degree of alignment (green dashed curve) computed using the optimized polarizability tensor parameters with the reference signal represented by the red solid curve (mean) and with 10 \% noise (blue symbols).
The optimized polarizability tensor parameters are $7.234~(-0.436), 8.028~(+0.268), 2.665~(+0.105), 0.985~(+0.135), 0.329~(-0.321)$. The values in parentheses are the deviations from the reference values (c.f., Table V). }
\end{figure}


\subsection{Effect of kernel complexity}

To build a GP model, one assumes an analytical form for the kernel function $k(\bm x, \bm x')$ with some unknown parameters.
The mathematical form of the kernel defines the GP model.
The choice of the kernel function is not unique.
As covariances are expected to decrease with the distance in the input space, one typically assumes a kernel function that decays with $|\bm x - \bm x'|$.  For example, in addition to the function (\ref{eqn:k_MAT}), the following functions are often used for GP models  \cite{mitchell1990existence, cressie1993statistics, stein1999interpolation}:

\begin{eqnarray}
k({\bm x}, {{\bm x}'})  =  {\bm x}^\top {{\bm x}'}& ~~~{\rm Linear~kernel}
\label{eqn:k_LIN}
\end{eqnarray}
\begin{eqnarray}
k({\bm x}, {{\bm x}'}) = \exp \left(-\frac{1}{2}r^2({\bm x}, {{\bm x}'})\right)&~~~~{\rm RBF~kernel}
\label{eqn:k_RBF}
\end{eqnarray}
\begin{eqnarray}
k({\bm x}, {{\bm x}'})  = \left ( 1 + \frac{|{\bm x}- {{\bm x}'}|^2}{2\alpha\ell^2} \right )^{-\alpha}&~~~~{\rm RQ~kernel},
\label{eqn:k_RQ}
\end{eqnarray}
where $r$ is the same as defined after Eq. (\ref{eqn:k_MAT}), $\alpha$ and $l$ are free parameters, and `RBF' and `RQ' are the abbreviations for `radial basis function' and `rational quadratic', respectively.

GP models with any kernel provide accurate interpolation in the limit of infinite training data and, consequently, BO with any GP model must converge to the target in the limit of an infinite number of iterations. Therefore, BO is typically performed with GP models trained using one of the kernel functions (\ref{eqn:k_MAT}) or (\ref{eqn:k_LIN}) - (\ref{eqn:k_RQ}).
However, the choice of the kernel function may affect the convergence of BO. Therefore, if the iterations are expensive, as is the case for quantum dynamics applications, the choice of the kernel function may be critically important.

The predictive power of GP models with the same number of parameters can be quantified by the magnitude of the log-likelihood function (\ref{log-likelihood-explicit}).
 However, in order to compare two GP models with kernel functions of different complexity, it is more suitable to use the Bayesian information criterion \cite{bic} defined as
\begin{eqnarray}
{\rm BIC} = \log {\cal L} - \frac{1}{2} {\cal M} \log n,
\label{BIC}
\end{eqnarray}
where $\cal M$ is the number of parameters in the kernel function and $n$ is the number of training points. The second term in Eq. (\ref{BIC}) penalizes kernels with more parameters.

As demonstrated previously \cite{extrapolation-1,extrapolation-2}, the magnitude of the BIC can be used as a model selection criterion to enhance the prediction power of GP models. The algorithm introduced in \cite{extrapolation-1,extrapolation-2} and used for physics applications in \cite{extrapolation-3,jun-dai}, builds composite kernels from the simple kernels (\ref{eqn:k_MAT}) and (\ref{eqn:k_LIN}) - (\ref{eqn:k_RQ}) as follows.
The kernel selection approach begins by training separately four GP models with each of the kernel functions (\ref{eqn:k_MAT}) and (\ref{eqn:k_LIN}) - (\ref{eqn:k_RQ}), denoted hereafter as $k_i$.
 The BIC (\ref{BIC}) is computed for each of the GP models with different kernel $k_i$ and the kernel function of the model with the largest BIC is selected as the preferred kernel $k_0 = k_i$.  The kernel $k_0$ is then combined with each of the four kernels (\ref{eqn:k_MAT}) and (\ref{eqn:k_LIN}) - (\ref{eqn:k_RQ}) by forming linear combinations $c_i k_0 + c_j k_j$ and products $c_j k_0 \times k_j$, thus leading to eight new functions. Eight new GP models are trained with each of these kernel functions by optimizing both the kernel parameters and the coefficients $c_i$ and $c_j$. The kernel of the model with the largest BIC is selected as the new preferred kernel $k_0$ and the procedure is iterated.

This process increases the complexity of the kernel function, making GP models more accurate. Here, we use this algorithm to increase the complexity of the GP models of the initial distribution of quantum dynamics calculations before BO, that is the model of the calculations to the left of the vertical broken line in Figure \ref{figure-1}. The kernel functions selected by the largest magnitude of the BIC are then fixed and used throughout the subsequent BO iterations.
Note that this kernel selection process does not require additional quantum dynamics calculations.

\begin{figure}
\centering
\includegraphics[scale=0.4]{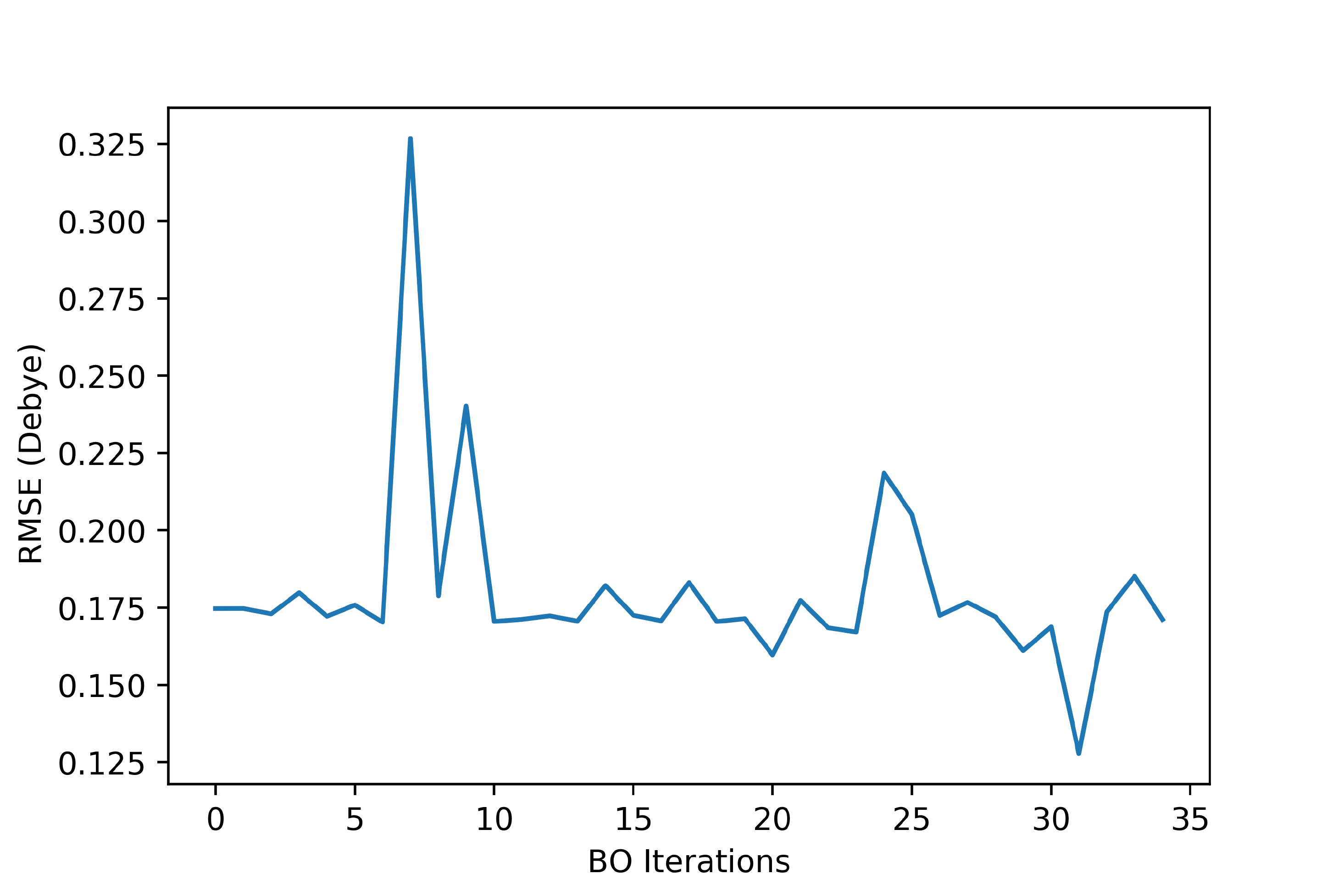}
\includegraphics[scale=0.4]{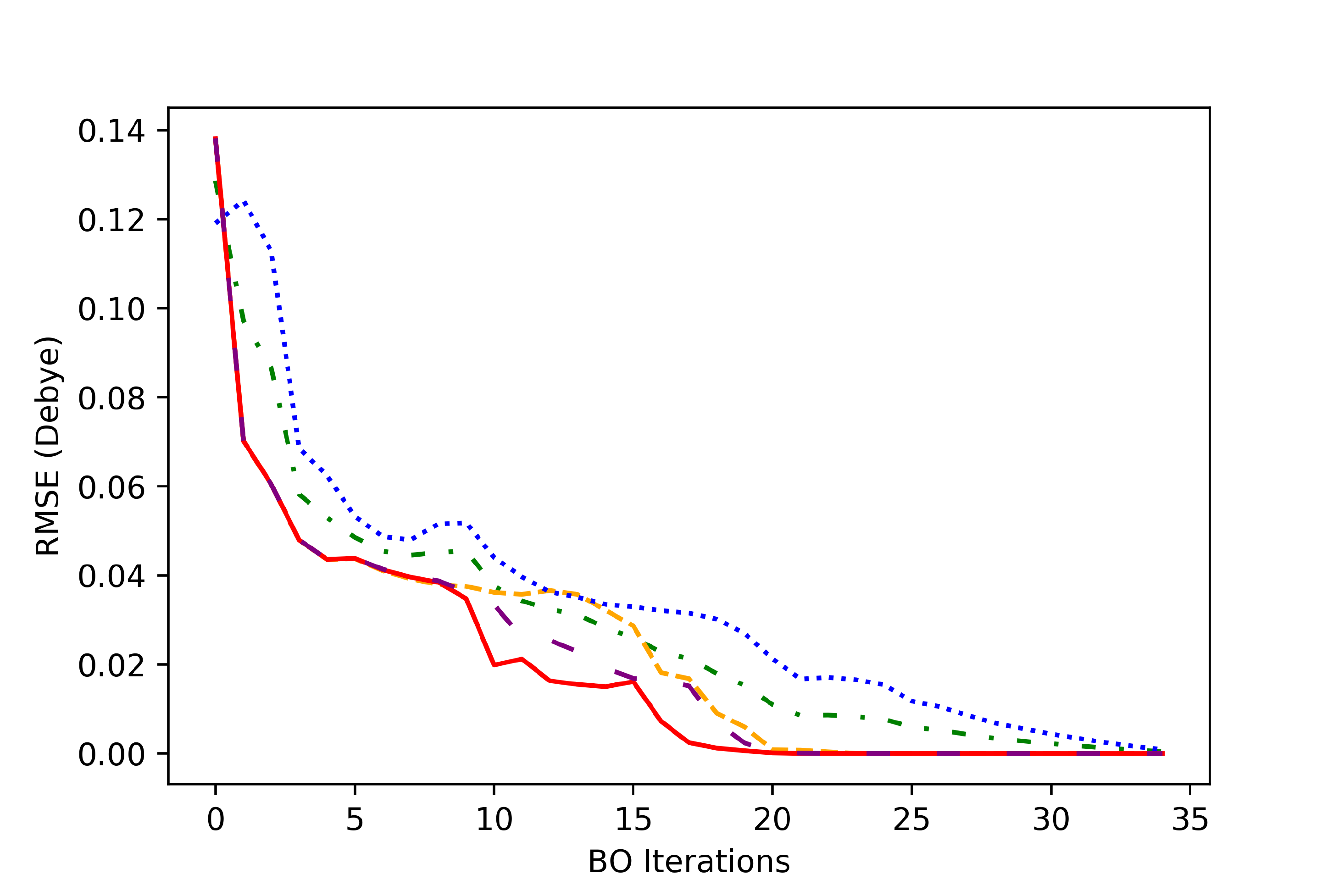}
\caption{Effect of kernel complexity on BO. Upper panel: BO with GPs using the linear kernel (\ref{eqn:k_LIN}); the value of BIC for the initial model with this kernel is 16142;
Lower panel (in the order of increasing BIC): BO with GPs using the Mat\'{e}rn kernel (blue dotted curve, BIC = 61269), RBF kernel (green dot-dashed curve, BIC = 65459), rational quadratic kernel (orange dashed, BIC = 75453),
a linear combination of RQ, Mat\'{e}rn and RBF (purple spaced dashed curve, BIC = 75465), and a linear combination of RQ and Mat\'{e}rn (red solid curve, BIC = 75470).
}
\label{figure-4}
\end{figure}

Figure \ref{figure-4} illustrates the effect of the kernel complexity on the convergence of BO (vector model) for the case of PPO with five variable polarizability tensor parameters. As can be seen, BO with the kernel functions corresponding to larger values of the BIC requires fewer iterations to converge. We have verified the generality of this result by repeating the calculations with different initial conditions.
It is important to note that the numerical effort to train GP models for inverse problems in quantum dynamics is a negligibly small fraction of the computation time required for solving the dynamics problem. Therefore, the algorithm described here substantially decreases the total computation time by decreasing the number of BO iterations required for convergence.

One can envision an extension of the present algorithm that re-optimizes the kernel complexity at each iteration of BO. Because the iterations are bottlenecked by the quantum dynamics calculations, this does not increase the computation time for inverse quantum problems.
We have tested this algorithm and found that optimizing kernels to increase the BIC of the GP models at each iteration does not affect significantly the total number of iterations required for convergence for the problems considered here. However, other problems may benefit from adjusting the kernels to yield Bayesian models with the largest BIC at each iteration.

\section{Conclusion}



We have demonstrated and compared three algorithms for inverse problems in time-dependent quantum dynamics based on Bayesian optimization of feedback loops.
The first algorithm is the straightforward application of BO designed to minimize a scalar function embodying the difference between predictions of the Schr\"{o}dinger equation and a reference observable.
This algorithm can be based on any Bayesian ML model and use one of several scalar functions to quantify the departure of quantum predictions from reference data, including the RMSE (\ref{target-rmse}) or Eq. (\ref{cos-sim}).
However, the convergence of BO depends on the underlying ML model.
Because each iteration of feedback loops requires numerical integration of the nuclear Schr\"{o}dinger equation, which is time-consuming, it is critically important to develop methods converging feedback loops with as few iterations as possible.

We have illustrated that the convergence of feedback loops is significantly accelerated when the time variable is discretized and the ML model used for BO is trained to reproduce the target observable at different time instances simultaneously.
This algorithm relies on Gaussian processes with multiple outputs and takes advantage of the model prediction and prediction uncertainties at different time instances. In the present work, the GP models exploited correlations between inputs into the models, but not between outputs. This algorithm can thus be further improved by also accounting for correlations between the outputs of a multi-output GP. However, this would make the acquisition function used for BO more complex than Eq. (\ref{aq-vector}).


Finally, we have illustrated that the convergence of feedback loops is further accelerated by increasing the kernel complexity of the GP models used for BO.
We proposed an approach that builds up the complexity of the GP models using a few training points before BO is initialized.
The kernel complexity is increased using a greedy search algorithm with the Bayesian information criterion for the model selection, as was previously done to enhance
pattern-recognition accuracy of GP models \cite{extrapolation-1,extrapolation-2, extrapolation-3}.
BO is then carried out with the complex kernels for the GP models thus obtained. We have shown that a combination of the vector approach based on multi-output GPs and
the method using GP models with complex kernels reduces the number of BO iterations by more than a factor of 3 compared to BO based on scalar GP models with simple kernels.

To illustrate these algorithms, we considered two-dimensional and five-dimensional inverse problems with several different observables.
We have shown that the convergence of BO is different for different observables, while the final results are the same.  We have also illustrated that BO is readily suited for solving inverse problems
using observables with Gaussian noise.
We note that the methods demonstrated in this work do not use explicitly any information on the evolution of the Schr\"{o}dinger equation solutions with the Hamiltonian parameters, yet the five-dimensional Hamiltonian parameter space
for problems considered here can be explored with as few as 30 quantum calculations.

It is important to emphasize that these conclusions are based on the empirical evidence obtained here for specific problems. However, the main qualitative conclusions regarding the acceleration of BO are expected to be general. To illustrate this, we note that the convergence of BO depends on the accuracy of the underlying GPs. Using the vector approach as described in this work effectively enhances inference at each iteration of BO by making
the acquisition function more efficient compared to those based on a single GP.
This is consistent with the conclusions in \cite{norm-minimization}.
Similarly, using composite kernels for GPs as described in this work, enhances the predictive power of GP for each iteration of BO.
We thus conclude that formulating inverse problems in quantum dynamics as a vector estimation problem for BO with composite kernels must generally reduce the total number of quantum calculations required to explore the Hamiltonian parameter space.

It is instructive to consider the extension of the present approach to problems with more dimensions. While specific applications may be affected by particular details, it is possible to make a few general observations.
The extension of the present approach to high-dimensional problems is limited by two factors: the numerical difficulty of the quantum dynamics calculations and the numerical complexity of building GPs for BO.
Training a GP by $n$ training points without approximations involves inverting an $n \times n$ matrix, which scales with $n$ as ${\cal O}(n^3)$.  The number of training points $n$ for accurate GP regression of a ${\cal D}$-dimensional problem can be estimated to be between $10 \times {\cal D}$ and
$20 \times {\cal D}$ \cite{sample-size}. As demonstrated in the present work, the kernel optimization algorithm used in conjunction with BO reduces this scaling to below the lower end of this estimate.
The numerical difficulty of training GP models should, therefore, become a significant factor only for problems with $\gtrsim 20$ -- $30$ unknown parameters. Kernel optimization becomes more difficult for higher-dimensional problems. However, we have recently implemented the kernel optimization algorithm used here for a 51-dimensional problem \cite{hiroki}. For high-dimensional problems, training of GP models can also be accelerated by well-controlled approximations, such as data sparsification \cite{sparse-0,sparse-1,sparse-2,sparse-3}.
The extension of the present approach to problems with up to 30 parameters is thus expected to be limited by the numerical difficulty of the quantum dynamics calculations.
Given our present results and the scaling analysis \cite{sample-size}, we estimate that a 30-dimensional problem will require between 180 and 600 quantum dynamics calculations, with kernel optimization reducing the problem to the lower end of the estimate.

We also would like to point out that the present approach is not restricted to a particular quantum dynamics simulation method and can be used for open as well as closed systems.
Quantum calculations for complex systems must rely on approximations. If quantum dynamics approximations lead to unknown uncertainties, the error of the quantum results will be absorbed into the Hamiltonian parameters determined by BO.
This is clearly undesirable.
To overcome this problem, one may use Bayesian model calibration \cite{BMC} to correct the results of quantum dynamics calculations either by isolated rigorous quantum results \cite{jie-prl} or by experimental data.
As shown in \cite{BMC-dynamics}, this can be achieved by using GP models with multiple outputs representing approximate and rigorous results. Such GP models are trained to learn correlations both between inputs and between outputs.
An output of such GP used for BO as in the present work should yield Hamiltonian parameters accounting for the error of the quantum dynamics approximation.

\clearpage

\newpage

\section*{Data availability}
The data that support the findings of this study are available from the corresponding author upon reasonable request.

\section*{Acknowledgments}
The authors appreciate valuable insights from Valery Milner, Sharly Fleischer and Dina Rosenberg on experimental aspects of measuring the laser-induced alignment and orientation of SO$_2$ and PPO molecules. This work was supported by NSERC of Canada and by Israel Science Foundation (Grant No. 746/15). I. A. acknowledges the support as the Patricia Elman Bildner Professorial Chair. This research was made possible in part by the historic generosity of the Harold Perlman Family.

\clearpage
\newpage



\end{document}